\documentclass[twocolumn]{revtex4}
\usepackage{graphicx}

\begin{document}

\title{Visualizing Free Energy Landscapes for Four Hard Disks}

\author{Eric R. Weeks$^*$}
\affiliation{Department of Physics, Emory University,
Atlanta, GA 30322, USA}
\email{erweeks@emory.edu}
\author{Keely Criddle}
\affiliation{Department of Physics, Emory University,
Atlanta, GA 30322, USA}

\date{\today}

\begin{abstract}
We present a simple model system with four hard disks moving
in a circular region for which free energy landscapes can be
directly calculated and visualized in two and three dimensions.
We construct several energy landscapes for our system and explore
the strengths and limitations of each in terms of understanding
system dynamics, in particular the relationship between state transitions
and free energy barriers.  We also demonstrate the importance of
distinguishing between system dynamics in real space and those
in landscape coordinates, and show that care must be taken to
appropriately combine dynamics with barrier properties to understand
the transition rates.  This simple model provides an intuitive
way to understand free energy landscapes, and illustrates the
benefits free energy landscapes can have over potential energy
landscapes.
\end{abstract}


\maketitle

%
%
%
%
%
%
%

\section{Introduction}
\label{intro}

Energy landscapes are of widespread utility for science,
relevant for understanding condensed matter systems
\cite{goldstein69,sciortino05,desouza08}, protein folding
\cite{bryngelson95,berry97,joseph17}, chemical reactions
\cite{marcelin1914,awasthi19}, optimization problems
\cite{krzakala07}, and even machine learning
\cite{ballard17}.  The first proposal related to
energy landscapes dates back to Ren{\'e} Marcelin, a French
physical chemist who in 1914 proposed understanding chemical
kinetics in terms of the Lagrangian coordinates describing atomic
motions \cite{marcelin1914,laidler85}.  Marcelin's idea was
``mouvement des points repr{\'e}sentatifs dans l'espace {\`a}
$2N$ dimensions'' -- ``movement of representative points in
$2N$-dimensional space,'' where $N$ is the number of Lagrangian
generalized coordinate pairs.  More modern descriptions of energy landscapes
date from 1969, when Martin Goldstein re-introduced the concept
\cite{goldstein69}.  Goldstein considered a situation with $N$
particles in a three-dimensional space, with some potential
energy $U$ of interaction between the particles.  To quote,
``When I speak of the potential energy surface I refer to $U$
plotted as a function of $3N$ atomic coordinates in a $3N+1$
dimensional space.''  That is, $U$ is a function of the $x,y,$
and $z$ positions of all $N$ particles (a total of $3N$ numbers),
so graphing $U$ forms a surface in this very high dimensional space.

\begin{figure}
\includegraphics[width=8cm]{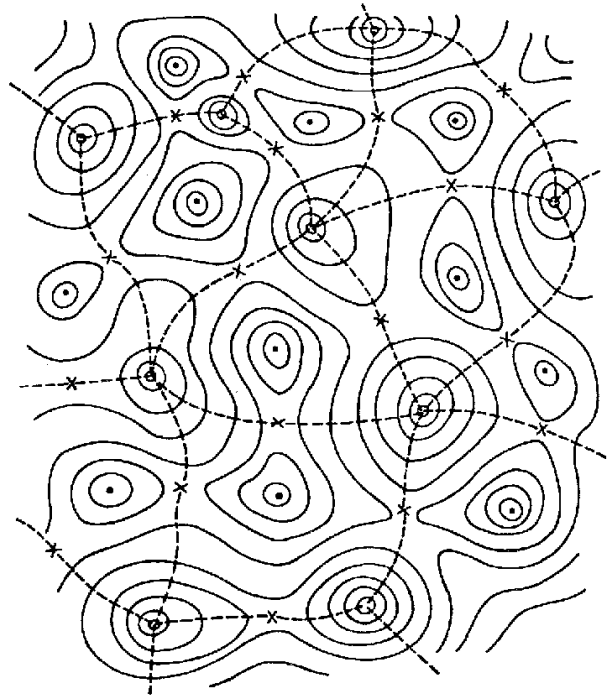}
\caption{
Sketch of a potential energy surface as a function of two
coordinates.  The solid lines are contours of constant $U$.  The
dashed lines separate local minimal (solid circles) and connect at
local maxima.  Saddle points are marked with $\times$.  From
Stillinger and Weber, {\it Science} {\bf 225}, 983-989 (1984)
\cite{stillinger84}.  Reprinted with permission from AAAS.
}
\label{stillinger_one}
\end{figure}

Picturing this high dimensional surface is of course challenging.  
The first picture the authors are aware of was published by
Stillinger and Weber in 1984 \cite{stillinger84}, and is shown in
Fig.~\ref{stillinger_one}.  Here the surface is represented as
the height as a function of two coordinates, giving rise to the
terminology of calling this the ``energy landscape.''  The solid
lines are contours of constant $U$.  The dashed
lines enclose local minima of the surface; the nodes where the
dashed lines connect are local maxima.  $\times$'s mark saddle points
between local minima.  For a thermal system, particles can
transition from one configuration to another by a thermal
fluctuation that carries them over a dashed line, perhaps
crossing near a saddle point.  The
transition in the $3N$-dimensional space encodes the appropriate
changes of the coordinates in real space of the $N$ particles.
Stillinger and Weber were interested in
studying liquids, so accordingly the disordered appearance of
Fig.~\ref{stillinger_one} represents the complex dependence of
$U$ on the amorphous liquid structure.

\begin{figure}
\includegraphics[width=8cm]{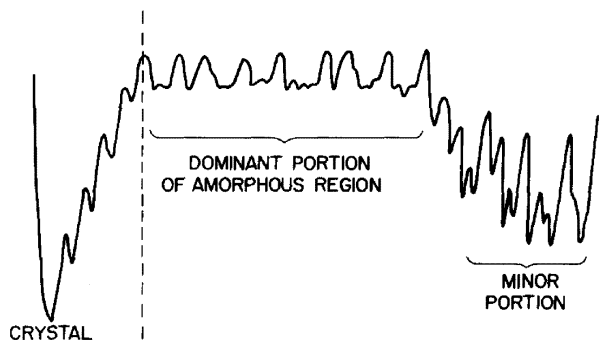}
\caption{
Sketch of a potential energy surface as a function of one
coordinate.  This particular sketch represents different possible
characters of the phase space, including a crystalline state
(left) and amorphous states (middle and right).  Reprinted from
F.~H.~Stillinger, {\it J. Chem. Phys.} {\bf 88}, 7818 (1988)
\cite{stillinger88}, with the permission of AIP Publishing.
}
\label{stillinger_five}
\end{figure}

A few years later, the challenge of drawing a surface as
a function of $3N$ coordinates was further simplified to a
curve in 1 dimension; a 1988 example from Stillinger is shown in
Fig.~\ref{stillinger_five} \cite{stillinger88}.
Here the lowest energy state corresponds to a crystal -- the
ground state.  In contrast, there are many disordered glassy
states and they have higher potential energy.  The states at the
right represent regions of phase space that correspond to
particular configurations of the $N$ atoms with slightly lower
$U$, but nonetheless which are amorphous and thus far from the
crystal configuration which minimizes $U$.  Later work
generalized this type of sketch to be more random in appearance
(for example, Ref.~\cite{stillinger95}), with the general
understanding that this one-dimensional landscape sketch
is supposed to convey a complex high-dimensional surface.

A conceptual simplification comes from considering the $N$
objects not to be atoms but rather hard spheres.  Hard spheres
have no attractive interaction, and repel each other if they
touch.  The potential energy $U$ for hard spheres is zero
if they do not overlap, and infinite if
they do overlap.  This system can be ordered into a crystal or
disordered like a liquid or glass, and so has interesting phase behavior
\cite{alder57,wood57,bernal64,widom67,pusey86}.  In this
situation, there are still $3N$ coordinates.  As a function of
these coordinates, $U$ is either zero or infinite, with
$U=\infty$ representing forbidden configurations where two or
more particles overlap.  Now transitions between states no longer
require crossing saddle points where $U$ is slightly higher;
rather, transitions between states require passing through
entropic bottlenecks \cite{zhou91,hunter12pre}.  One can think of a free
energy, $F = U - TS$, 
where $T$ is the absolute temperature and $S$ is the entropy.
States with high $S$ (thus low $F$) 
correspond to common configurations of the hard spheres, and
states with low $S$ correspond to rare configurations.
One would have to understand 
what would be meant by a common or rare configuration
of the $3N$ coordinates, and it is not obvious how a sketch
of $F$ would differ (if at all) from something like
Figs.~\ref{stillinger_one} or \ref{stillinger_five}.  
While hard spheres are a conceptual
simplification, it does not necessarily make understanding the
free energy landscape any simpler.

The earliest mention of free energy landscapes that we are
aware of was Hill and Eisenberg in 1976, who considered ``free
energy surfaces'' in myosin-actin-ATP systems \cite{hill76}.
They projected the behavior of the system down to a few important
coordinates.  For example, at one point they discuss the free
energy of a molecule based on one coordinate, ``statistically
averaged over all possible configurations of all solvent molecules
and over the rotational coordinates of the ligand.''  Later work
was more explicit about the landscape analogy; for example
Bryngelson and Wolynes in 1987, also considering a biological
system (protein folding), discuss ideas such as moving downhill
to a local minimum of the free energy.  Other work in the 1980's
and 1990's talked about free energy landscapes but gloss over
the difference between that and a potential energy landscape;
these authors mainly consider free energy landscapes as they had
methods to calculate the free energy as a function of coordinates
\cite{soukoulis82,paine85,fontana91}.  A common approach for free
energy is to consider just one or two ``reaction coordinates'' or
``order parameters''
that describe a behavior of interest, such as protein folding or
a chemical reaction \cite{wales03}.  A recent review explicitly
addressing both potential energy and free energy landscapes notes
that for the latter, one constructs a free energy landscape by
``by averaging over most of the coordinates'' \cite{wales06}.
The main benefit of free energy landscapes is to focus attention
on a small number of meaningful coordinates.

\begin{figure}
\includegraphics[width=5cm]{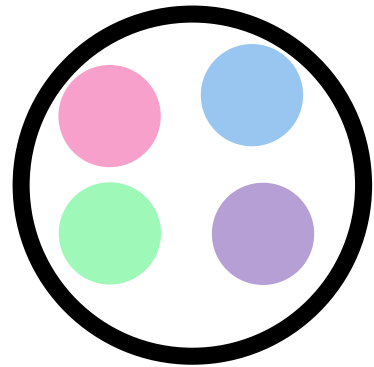}
\caption{
(Color online) Illustration of our model system, with four hard
disks confined to a two dimensional circular region.  The disks
have radius equal to 1, and the circular region has a radius
$3+\epsilon$; in this sketch $\epsilon = 0.3$.  
}
\label{demo}
\end{figure}

In this paper we present a model system using four hard disks
which has nontrivial dynamics and a nontrivial free energy
landscape.  The potential energy landscape as a function of the
$2N = 8$ coordinates can be usefully projected down to three or
even two dimensions so that a free landscape can be directly
visualized, rather than needing a conceptual sketch.  We use
this model system to illustrate several ideas about free energy
landscapes.  For example, a key point is that this projection
operation is not unique:  there are multiple possible ways to
visualize the free energy landscape, with varying utility.
We verify that similar results are obtained for diffusive
dynamics and ballistic dynamics.  Our model is in the spirit of
other simple models involving very small numbers of particles
\cite{hunter12pre,giarritta94,speedy94,bowles99,bowles06,ashwin09,carlsson12,barnettjones13,hinow14,godfrey14,godfrey15,robinson16}.

\section{Dynamics of the model system}

\begin{figure}
\includegraphics[width=8cm]{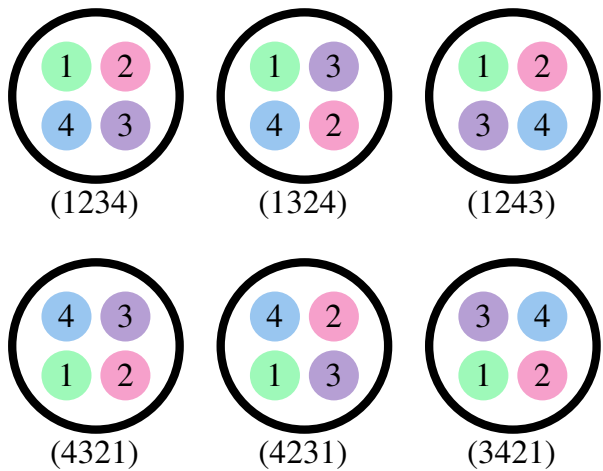}
\caption{(Color online)
The six distinct ``equilibrium'' states, labeled by the order of
the disks in a clockwise direction.  The bottom row of states are
mirror images of the top row.  
At any given moment it is unlikely that the four particles are
exactly arranged in a square, but rather these states should be
understood as the situations where the order of the four particles
is well defined.  Likewise, rotations of these states are considered
equivalent; only the order matters.  For example, the state shown
in Fig.~\ref{demo} is (1243).
}
\label{sixstates}
\end{figure}

\begin{figure}
\includegraphics[width=8cm]{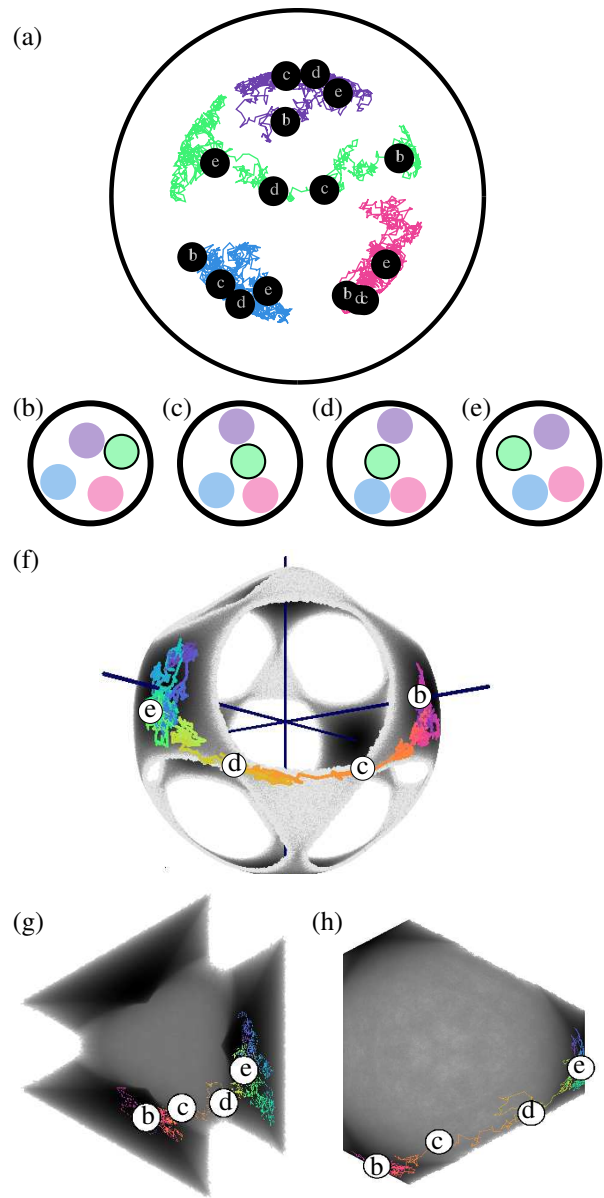}
\caption{(Color online)
(a) Real space trajectories during a transition
with system size $\epsilon = 0.18$ and using diffusive
dynamics.  The letters
correspond to the snapshots of the system shown in panels (b-e).
In this transition $(1243) \rightarrow (1324)$, the outlined disk
moves through the middle.  The total time pictured is $10\tau_D$,
and steps are drawn spaced by $\Delta t = 0.01\tau_D$.
(f) Free energy landscape in the variables $(c_1,c_2,c_3)$.
(g) Free energy landscape in the variables $(d_1,d_2)$.  (h) Free
energy landscape in the variables $(u_1,u_2)$.  
In panels (f,g,h), 
the positions corresponding to snapshots (b) through (e) are marked.
}
\label{bigfig}
\end{figure}

\begin{figure}[thb]
\includegraphics[width=8cm]{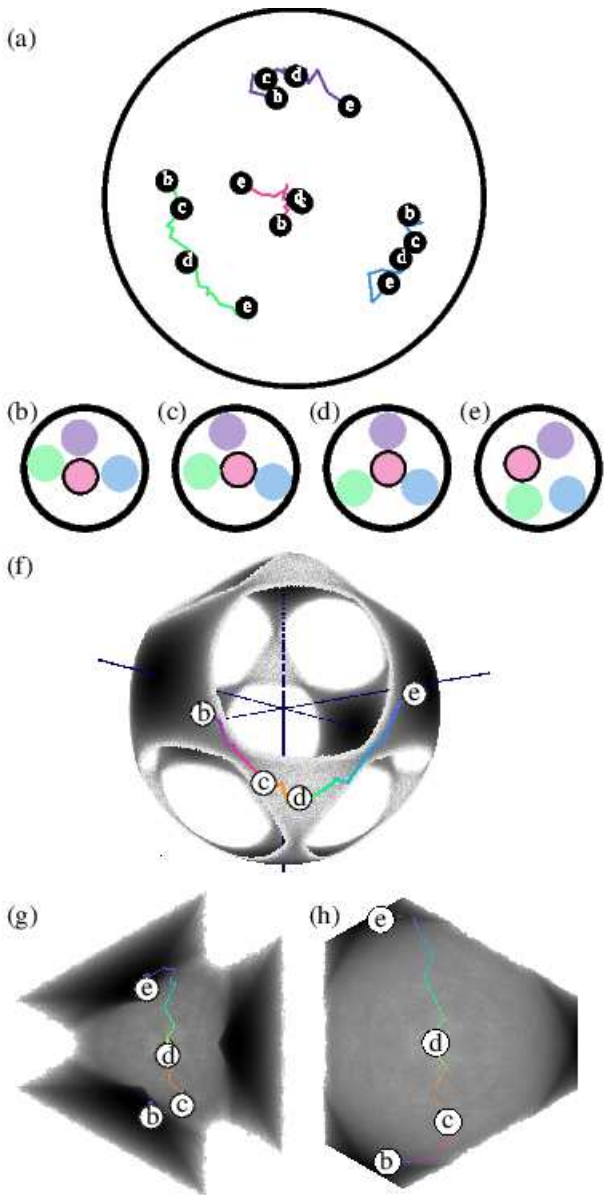}
\caption{(Color online)
(a) Real space trajectories during a transition
with system size $\epsilon = 0.20$ and using ballistic
dynamics.  Letters
correspond to the snapshots shown in panels (b-e).
In this transition $(3421) \rightarrow (1234)$, the outlined disk
moves through the middle.  The total time pictured is $4\tau_v$.
(f) Free energy landscape in the variables $(c_1,c_2,c_3)$.
(g) Free energy landscape in the variables $(d_1,d_2)$.  (h) Free
energy landscape in the variables $(u_1,u_2)$.  
In panels (f,g,h), the 
positions corresponding to snapshots (b) through (e) are marked.
Note that the phase space trajectory in (f) has been rotated;
in particular this is
a different perspective than shown in Fig.~\ref{bigfig}(f).
}
\label{mbigfig}
\end{figure}

Figure \ref{demo} shows the model system, comprised of four hard disks
confined to a two-dimensional circular region; this is an
extension of a previous model with three hard disks
\cite{hunter12pre}.  We let the disks
move, subject to the constraint that no disk can overlap
another disk or overlap the boundary of the system.  The disks
are distinguishable, so there are six ``equilibrium'' states,
shown in Fig.~\ref{sixstates}.  Changing from one state to
another requires one of the disks to move through the middle of
the system so as to swap locations with one of its neighbors.
Examples of these swaps are shown in
Fig.~\ref{bigfig}(a-e) and Fig.~\ref{mbigfig}(a-e).  Swapping
locations requires a large enough system for this to occur:  three
disks must be able to align momentarily as the middle one passes
through the other two.  We define
the disks all to have radius 1, and then the minimum system size is
radius $R=3$.  A smaller $R$ is possible, but then no rearrangements can
occur.  A larger system makes rearrangements easier, so
accordingly we define the system radius to be $R = 3 + \epsilon$.
Letting $\epsilon \rightarrow 0$ results in behavior a bit like a
crystalline or glassy system, in that particles become unable to
rearrange, although they can still vibrate locally.

We use two simulation methods.  The first method approximates diffusive
motion for the particles, and will be used for most of the
results presented in this paper.  In this simulation, we consider
a small trial
move for a disk in a random direction.  This move is accepted if
the new position does not overlap any other disk or the boundary,
and otherwise is rejected.  A simulation time step occurs when
we have considered one trial move for each of the four disks
(picked in random order at each time step).  We choose a step size
of $L=10^{-2.5}$ so that most steps are accepted, and verify that
our results are insensitive to this choice.  With this choice,
the time it takes for a free particle to diffuse in the $x$ (or
$y$) coordinate a distance 1 is given by $\tau_D = 1/L^2 = 10^5$
time steps.  Accordingly, we define our time in units of $\tau_D$.
We run our simulations for $10^4 - 10^5 \tau_D$, long enough for at
least 20 rearrangements to occur, and often $100-1000$
rearrangements, depending
on $\epsilon$.  As will be seen, the smaller $\epsilon$ is the
longer it takes for a rearrangement to occur.

The second simulation method computes ballistic trajectories for
each disk using an event-driven computation.  For this
simulation, the four disks are initialized with velocities $v=1$
in random directions, but with the constraint that the total
angular momentum is zero.  We calculate the next time for each
possible collision (disk-disk or disk-wall) and advance the
positions of the four particles to the earliest collision.  The
velocity of the colliding particle(s) are updated conserving
energy and momentum.  Where these results are presented in this
paper, time is in units of $\tau_v = 1/v = 1$, the time it takes a
non-colliding particle to move a distance of 1 (the disk radius)
based on the initial velocity scale $v$; note that the
instantaneous velocity of the disks fluctuate due to the
collisions, albeit with the total kinetic energy constant.

There are many possible ways to project from the 8-dimensional
phase space in the disk coordinates down to lower dimensions.
One needs to map from the four original positions down to a
smaller number of coordinates.  We choose to use vector
operations.  Relative positions are describable
by vectors pointing from disks $i$ to disks $j$:
\begin{equation}
\vec{v}_{ij} = (x_j - x_i, y_j - y_i).
\label{vecv}
\end{equation}
With $i \ne j$, this is a set of six vectors (ignoring the
counterparts in opposite direction, that is, using $\vec{v}_{12}$
and not $\vec{v}_{21} = - \vec{v}_{12}$).  While many operations
could be done with these vectors to generate landscape coordinates,
it is easiest to consider working with pairs of vectors:  there
are 15 such pairs.  It also is useful to require each pair of
vectors to depend on the coordinates of all four particles: this
reduces the number of distinct pairs to 3.  That is, considering
the pair ($\vec{v}_{12}, \vec{v}_{13}$) is not desirable as it
tell us nothing about particle 4, whereas the pair ($\vec{v}_{12},
\vec{v}_{34}$) has some information about all four particles.
Finally, we will consider the two straightforward vector
operations to act on each pair of vectors:
the cross product and the dot product, which will each result in
a distinct free energy landscape.

We first consider the cross product, and will use this initially
to illustrate the system dynamics.
We compute the vector cross products:
\begin{eqnarray}
c_1 &=& (\vec{v}_{12} \times \vec{v}_{34}) \cdot \hat{z},\\
c_2 &=& (\vec{v}_{13} \times \vec{v}_{42}) \cdot
\hat{z},\nonumber\\
c_3 &=& (\vec{v}_{14} \times \vec{v}_{23}) \cdot
\hat{z},\nonumber
\label{cross}
\end{eqnarray}
where the final dot product with $\hat{z}$ ensures that the
$c$'s are scalars; $\hat{z}$ is the unit vector perpendicular to
the two-dimensional system.  
For the equilibrium configurations shown in
Fig.~\ref{sixstates}, the $c$'s are positive, negative, or roughly
zero depending on the arrangement of the four disks.
For example, if the disks are
arranged in a square of side length $s$, with the disks arranged
(1243), then $c_1 = +2s^2$, $c_2 = c_3 = 0$.  If the disks are
arranged in the opposite order (3421), then $c_1 = -2s^2$ and
$c_2 = c_3 = 0$.  Likewise $c_2$ and $c_3$ are each nonzero for
two opposite pairs of configurations, and zero for the other four.

%
%

\begin{figure}
\includegraphics[width=8cm]{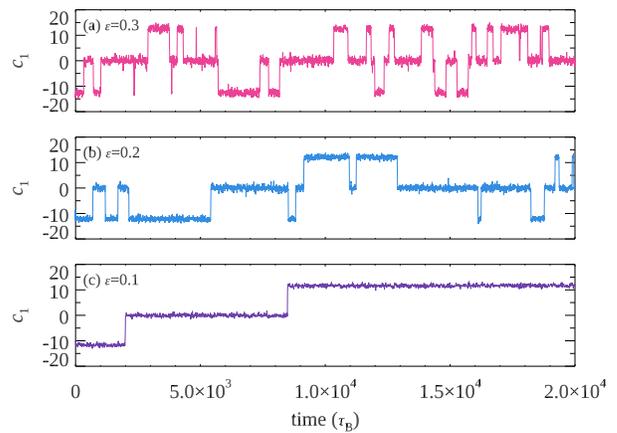}
\caption{(Color online)
Trajectories of the coordinate $c_1$ as a function of time, for
system sizes $\epsilon$ as indicated, using diffusive dynamics.  
Transitions indicate one
of the disks passing through the middle of the system.
Note that $c_1$ is zero for four of the six states shown
in Fig.~\ref{sixstates}, so
some transitions that keep $c_1=0$ are not apparent in the data,
but would be apparent in plots of $c_2$ and $c_3$.
}
\label{c1}
\end{figure}

\begin{figure}
\includegraphics[width=8cm]{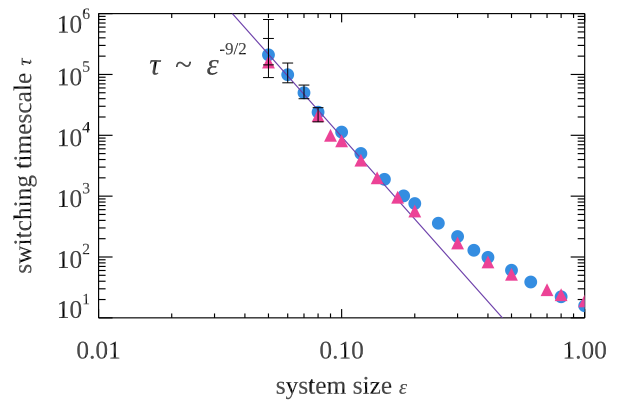}
\caption{(Color online)
The switching timescale $\tau$ as a function of the system size
$\epsilon$, for diffusive dynamics (circles, in units of
$\tau_D$) and ballistic dynamics (triangles, in units of
$\tau_v$).  As $\epsilon \rightarrow 0$ the timescale grows as
$\epsilon^{-9/2}$ as indicated by the solid line.  
The uncertainties are 95\% confidence intervals;
where not shown, the uncertainties are smaller
than the symbol size.
}
\label{lifetime}
\end{figure}

Before constructing a free energy landscape with these coordinates,
first consider the behavior of the system viewed through one of
these coordinates, shown in Fig.~\ref{c1}.  For a relatively large
system size [panel (a), $\epsilon = 0.3$], transitions happen
fairly frequently.  As the system size is decreased, panels (b)
and (c) show transitions happen less frequently.  This is because
there is less ability for the disks to find a configuration where
one disk passes through the middle of the system to swap places with
one of the others.  

Figure \ref{lifetime} shows the mean time $\tau$ between switching
states as a function of the system size $\epsilon$.  As $\epsilon
\rightarrow 0$ the switching time grows larger, confirming the
qualitative picture of Fig.~\ref{c1}.  It is perhaps a bit of
a coincidence that as a function of $\epsilon$ the magnitude of
$\tau$ is similar for diffusive dynamics (circles, in terms of
$\tau_D$) and ballistic dynamics (triangles, in terms of $\tau_v$).
The particular power law dependence $\tau \sim \epsilon^{-9/2}$
will be explained in Sec.~\ref{quantify}.

The behavior appears similar to a glass transition, in that the time
scale for rearrangement grows dramatically as $\epsilon \rightarrow
0$.  As a molten glass is cooled, its viscosity grows dramatically
-- which is to say, the time scale for internal rearrangements grows
dramatically \cite{sciortino05,ediger12,biroli13}.  The previous
three disk model (which inspired this four disk model) was designed
to capture the basic crowding that can lead to a glass transition
\cite{hunter12pre}.  The coordinated motion of the four disks
during a transition from one basic state to another conceptually
resembles what is seen in simulations of materials close to the
glass transition \cite{desouza08,donati98,doliwa00}.

\section{Free Energy Landscapes}

We wish to use the simulation data to map the free energy landscape.
In the original senses of Marcelin \cite{marcelin1914} and
Goldstein \cite{goldstein69}, the potential energy is
an 8-dimensional landscape as we have four disks each of
which is described by two coordinates $(x,y)$.  Some states are
allowed (states such that no disks overlap) and those all have
equal probability.  To generate a more interesting free energy
landscape, we must project the 8-dimensional description down
to lower dimensions, where we will see that states do not have
equal probability -- thus leading to an entropic penalty for some
states, and a nontrivial free energy landscape.

Before further choosing a projection, we first consider
what transitions between the states are possible.  In
Fig.~\ref{sixstates}, consider changing from one of the states
to another one.  For example, changing between (1234) to (1324)
requires swapping disks 2 and 3.  This can be done by having
disk 2 move to the middle of the system, and then swap places
with disk 3; or likewise disk 3 could be the one to move through
the middle.  Changing from (1234) to (4321) requires two such
swaps, as simultaneously swapping two disks across the diagonal
requires $R > 4$.  In fact, starting at any one of the states in
Fig.~\ref{sixstates}, there are four choices of adjacent particle
pairs that could be swapped, leading to four different new states.
The only state for which a direct transition is disallowed is
the mirror image state, which requires two particle pair swaps.
The easiest way to picture this is to have each state correspond
to a face of a cube.  Only transitions between adjacent cube faces
are allowed.

\begin{figure}
\includegraphics[width=8cm]{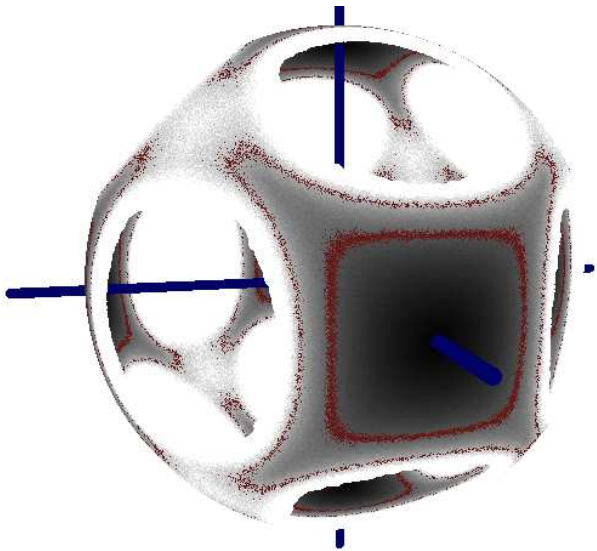}
\caption{(Color online)
Free energy landscape in the variables $c_1, c_2, c_3$ for system
size $\epsilon = 0.5$, projected onto a unit sphere.  The blue
lines indicate the axes ($c_1$, etc.) and the red points are
contours at $2k_BT$ (the square-shaped contour) and $4k_BT$ (the
stretched contour).  The black regions pierced by the axes
correspond to the lowest points in the landscape (states with
highest entropy).
}
\label{spherespace}
\end{figure}

The phase space of $c_1,c_2,c_3$ (Eqns.~\ref{cross}) has
the desired cubical symmetry for our free energy landscape.
To generate the landscape, we compile
a histogram of the microstates seen in the simulation, $\Omega(c_1,c_2,c_3)$.
The entropy is then $S = k_B \ln \Omega(c_1,c_2,c_3)$ using
Boltzmann's constant $k_B$.  The free
energy landscape is $F = U - TS = -TS$ (since $U=0$ must be true
for non-overlapping hard disks).  Equivalently, we can consider
$-\ln \Omega$ to be the free energy landscape in units of $k_B T$.
In this three-dimensional phase space, it turns out that states
near the origin are never seen (for $R<4$), so this phase space can be safely
projected onto the surface of a unit sphere.  This projection is
shown in Fig.~\ref{spherespace}, where the large squarish regions
correspond to the equilibrium states, and the tenuous connections
through the triangular corner regions show transition
paths between the equilibrium states.  The empty circular regions
correspond to the cube edges, which are configurations that would
cause disks to overlap and thus are forbidden.  

This is our first free energy landscape.  The colors in
Fig.~\ref{spherespace} indicate the height of the free energy
landscape, with darker colors being the minima corresponding to
the equilibrium states.  To change states, the system must undergo
a real-space rearrangement which corresponds to moving from a
cube face ``up'' the energy landscape to one of the corners,
and then back ``down'' to a different cube face; one such
trajectory is shown in Fig.~\ref{bigfig}(f).  Given that
this is mapped to the surface of a unit sphere, this landscape
is a function of only two (angular) coordinates, although given
the cubical symmetry it is perhaps more useful to think of this
as a function of the three $c$'s which have the proper symmetry.
Nonetheless, it is intriguing that the 8 original coordinates can
be usefully reduced down to two or three effective coordinates in
this free energy landscape.

We now consider a second free energy landscape.  
Three dot products can be defined as:
\begin{eqnarray}
\label{qeqn}
q_1 &=& \vec{v}_{12} \cdot \vec{v}_{34},\\
q_2 &=& \vec{v}_{13} \cdot \vec{v}_{42},\nonumber\\
q_3 &=& \vec{v}_{14} \cdot \vec{v}_{23}.\nonumber
\end{eqnarray}
With these choices, if the four disks are arranged at the
corners of a square of side length $s$ such as in Fig.~\ref{sixstates},
one of the
dot products will be zero and the other two will be $\pm s^2$.
However, states that are reversed are indistinguishable:  (1234)
is identical to (4321).  Thus, rather than six unique states
with cubical symmetry, there are three unique states with
triangular symmetry.  They are
$(q_1,q_2,q_3)=(s,-s,0); (-s, 0, s);$ and $(0,s,-s)$.
In three dimensions, these are the corners of an equilateral
triangle.  Given that these three points span a plane, we can
project the data onto a 2D plane by defining three mutually
perpendicular unit vectors:
\begin{eqnarray}
\label{dot}
\hat{d_1} &=& (+1, -1, 0) / \sqrt{2},\\
\hat{d_2} &=& (+1, +1, -2) / \sqrt{6},\nonumber\\
\hat{d_3} &=& (+1, +1, +1) / \sqrt{3}\nonumber
\end{eqnarray}
where $\hat{d_1}$ is directed toward the $(s,-s,0)$ location,
$\hat{d_2}$ is chosen to be in-plane and perpendicular to
$\hat{d_1}$, and $\hat{d_3} = \hat{d_1} \times \hat{d_2}$.
In the plane spanned by $\hat{d_1}$ and $\hat{d_2}$, we define
coordinates by $d_1 \equiv \hat{d_1} \cdot (q_1,q_2,q_3)$, $d_2
\equiv \hat{d_2} \cdot (q_1,q_2,q_3)$.  It turns out that $\hat{d_3}
\cdot (q_1,q_2,q_3) = 0$ which can be shown by putting in the
definitions of $q_1,q_2,q_3$ in terms of the original disk
positions.  This shows that the coordinates $(q_1,q_2,q_3)$
lie on a plane rather than filling a three-dimensional region.

\begin{figure}
\includegraphics[width=9cm]{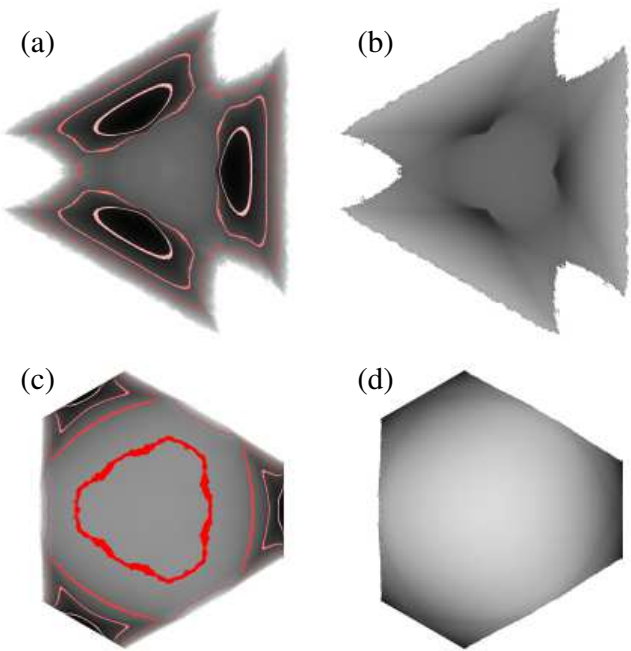}
\caption{
(a) Free energy landscape in $(d_1,d_2)$ coordinates for
$\epsilon=0.4$.  The landscape is bounded by the whitest points.
Within the landscape, darker regions are lower (more probable).  The
red contours are at $1k_BT,$ $3k_BT$, and $5k_BT$.  The center of
the landscape is a broad local maximum with a height of $4.9k_BT$.
(b) Map of the
effective diffusivity at each point in the landscape.  The
diffusivity in the center is 1.22 times greater than the
diffusivity in the three darker spots.  The diffusivity near the
long straight outer edges of the landscape is 1.48 times greater than the
diffusivity in the darker spots.  (c) Free energy landscape in
the $(u_1,u_2)$ coordinates for $\epsilon=0.4$.  The red contours
are at $1k_BT,$ $3k_BT$, $5k_BT$, and $7k_BT$.  The center of the
landscape has a height of $7.3k_BT$.  (d) Map of the effective
diffusivity at each point in the landscape.  The diffusivity in
the center is 2.8 times greater than the diffusivity at the dark
edges.
}
\label{diffuse}
\end{figure}

A visualization of the 2D $(d_1,d_2)$ free energy landscape is shown
in Fig.~\ref{diffuse}(a).  The dark regions are the equilibrium
states, and the brighter regions correspond to higher locations
on the free energy landscape (lower entropy, and thus the
unlikely transition regions).  While this representation
collapses the six equilibrium states into three minima,
nonetheless all transitions are seen in this free energy
landscape as landscape trajectories from one local minimum to
another one.  One such transition is shown in
Fig.~\ref{bigfig}(g).

One final free energy landscape that is useful to consider is formed
by defining the $q$ variables (Eqns.~\ref{qeqn}) using unit vectors;
that is, changing from $\vec{v}_{12}$ to $\hat{v}_{12}$.  The
landscape coordinates are then defined using the $\hat{d}$ vectors
given in Eqns.~\ref{dot}, leading to coordinates $(u_1, u_2)$ in
analogy with $(d_1,d_2)$.  The landscape for these coordinates is
shown in Fig.~\ref{diffuse}(c).  Figure
\ref{bigfig}(h) illustrates a trajectory through this landscape.

Crossings through the exact middle of the phase space [either the
$(d_1,d_2)$ or ($u_1,u_2$) phase space] correspond to an
unusual situation where the transition is equally likely to go to
any of three different equilibrium states, as shown
in Fig.~\ref{mbigfig}.  The disks near the edge
of the system become symmetrically placed around the disk in
the center, as shown in Fig.~\ref{mbigfig}(d).  This arrangement
allows the central disk to be equally likely to go to any of the
three possible equilibrium states.  In the cubical free energy
landscape of Fig.~\ref{spherespace}, this configuration
corresponds to the centers of the corners of the cube
[Fig.~\ref{mbigfig}(f)], where
transitions to any of the adjacent three faces is equally likely.

These three free energy landscape representations (the $c$ variables
using the cross products, the $d$ variables using the dot product,
and the $u$ variables using the dot product of unit vectors)
illustrate our first point about free energy landscapes:  
{\bf Multiple free energy landscapes can be constructed to represent the
same system.}  This point was also made in Ref.~\cite{hunter12pre},
which presented two different one-dimensional landscapes for
a system with three hard disks.  To an extent this observation
is trivial:  even the original coordinate system is arbitrary.
One could use Cartesian $(x,y)$ coordinates to describe the position
of each disk, or polar coordinates $(r,\theta)$.  It is reasonable
that likewise a free energy landscape could be described
by different coordinates.

\begin{figure}
\includegraphics[width=8cm]{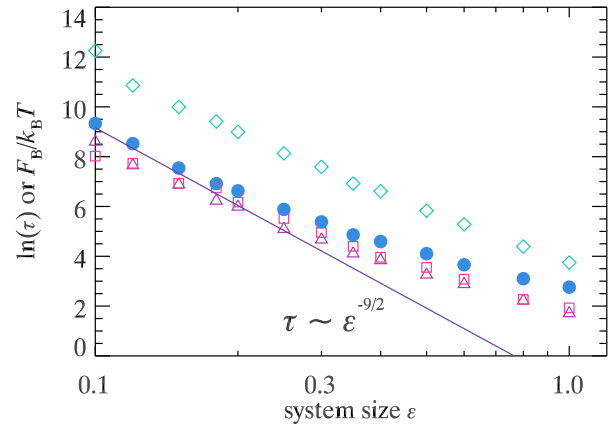}
\caption{(Color online)
The blue circles are $\ln(\tau)$, the log of the switching time
scale.  The straight line is the same as shown in
Fig.~\ref{lifetime}.  The diamonds, squares, and triangles indicate the free
energy barrier height for the $u$, $c$, and $d$ free energy landscape
variables, respectively.  Uncertainties for all points are
smaller than the symbol size.
}
\label{lifebarr}
\end{figure}

However, one fact is intriguing:  the different
representations [Figs.~\ref{spherespace} and \ref{diffuse}(a,c)] lead
to different free energy barrier heights!  The barrier heights
are plotted in Fig.~\ref{lifebarr}.  The free energy barrier
heights are similar for the cross
products landscape (squares) and the original dot
products landscape (triangles), and markedly higher for the unit
vector landscape (diamonds).  For the first two, it is clear that
$\ln(\tau) \sim F_B$, that is, the switching time scale grows
essentially exponentially with the barrier height.  (The 
deviation from this relationship at large $\epsilon$ is due
to the large system size where the disks require more time to
diffuse across the system in order to have a transition
\cite{hunter12pre}.  That
is, as $R$ gets large, the disks spend more time farther
apart from one another, and thus the switching time is no longer
dominated by the free energy barrier.)  The 
free energy barrier from $(u_1,u_2)$ is not only consistently
larger than the other two, but also grows faster as $\epsilon
\rightarrow 0$ than the other two barriers.

\begin{figure}
\includegraphics[width=8cm]{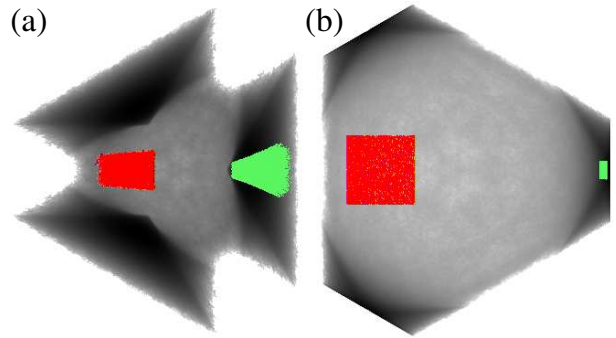}
\caption{(Color online)
Free energy landscapes with $\epsilon=0.2$ for (a) the $(d_1,d_2)$
variables and (b) the $(u_1,u_2)$ variables.  Equivalent points
in the two phase spaces are marked by the red regions (left side
of both images) and the green regions (right side of both
images).
}
\label{omega}
\end{figure}

This brings us to our second point about free energy landscapes:
{\bf The nonlinear mapping used to create free energy landscapes can
distort the free energy barrier height.}  This is especially true
when comparing different free energy landscapes.  To understand
this, consider the mapping from the original 8-dimensional space
to a projected free energy landscape.  There is some region of
the original 8-dimensional space corresponding to the transition
states between equilibria with size $\Omega_t$, and another region
corresponding to the equilibria states of size $\Omega_e$.  Figure
\ref{omega} shows that these do not map to equivalent proportions
of different landscapes:  $\Omega_t$ takes up a large portion of
the $(u_1,u_2)$ landscape and $\Omega_e$ takes up a small portion,
as seen in Fig.~\ref{omega}(b), large red square and small green
rectangle, respectively.  However, they map onto nearly equivalent
areas in the $(d_1,d_2)$ landscape, as seen in Fig.~\ref{omega}(a).
For any given small region in the central transition region of
Fig.~\ref{omega}(b), there are fewer microstates -- the density of
microstates per unit area is lower -- and thus the entropic barrier
$\sim -\ln \Omega$ is higher.  Also important is that the density of
microstates in the equilibrium region is higher [comparing the
green points of Fig.~\ref{omega}(a,b)], thus increasing the
entropy associated with those common states for the $(u_1,u2)$
landscape, which further increases the entropic barrier for the
rare transition states.

To be clear, all of these projections are valid and are free
energy landscapes for the same system -- and the switching time
between equilibrium states cannot depend on how we represent the
free energy landscape.  To understand how the switching time is
independent of the free energy landscape, we need to understand
how diffusive motion occurs on each landscape.
As pointed out by Frenkel \cite{frenkel13}, 
the switching time between two minima is a product of the
barrier height and the time it takes to move across the barrier.
The latter is based on the barrier width and the diffusion rate
across that region.
As seen in Fig.~\ref{diffuse}(a,c) the barrier is much wider
for the $(u_1,u_2)$ coordinates as well as taller -- but also,
the system diffuses through this region more quickly.  This can
be seen by comparing the segment sizes between panels (g) and (h)
of Fig.~\ref{bigfig}.  Each segment corresponds to 1000 simulation
time steps, and these correspond to larger relative distances
in phase space in Fig.~\ref{bigfig}(h).  The combined influence
of barrier height, barrier width, and diffusion rate through the
transition region are such that the time scale for a transition
is the same for all free energy landscapes -- as it must be.  As
it happens, the $(c_1,c_2,c_3)$ and $(d_1,d_2)$ landscapes have
essentially unchanging crossing attempt rates as $\epsilon
\rightarrow 0$ and thus $\ln(\tau) \sim F_{\rm B}$ stays true.
This suggests that these two landscape representations are more
``useful'' to a physicist.  To be clear, this discussion has
focused on diffusion; similar comments would apply to the
ballistic dynamics, where a constant real-space velocity yields
different rates of transport in different landscapes.

Effective diffusion rates can vary between landscapes, but the
story of diffusion on a free energy landscape is more complicated
than that.
A third point about free energy landscapes is {\bf diffusion
rates on the free energy landscape can vary spatially.}  This is
demonstrated in Fig.~\ref{diffuse}(b,d), which shows the the
effective diffusivity at each point in the landscapes corresponding
to panels (a,c).  The effective diffusivity was determined by
simulating the system for a long time ($>100$ transitions).
At each time we save the position that the system is at in the
landscape, and then measure how much that position changes in the
next simulation time step.  The effective diffusion is measured
as the mean square landscape displacement, as a function of the
initial coordinate.  Recall that at each time step, disks
can move a distance $10^{-2.5}$ in any direction in real space:
this is always the same no matter where the disk is, with the
exception of configurations where that movement is disallowed so
that overlaps are avoided.  In the landscape, however, the motion
depends on the transformation from the 8 real space coordinates
into the landscape coordinates, and this is nonlinear.  The real
space motions result in a smaller or larger movement within the
landscape depending on the position, as Fig.~\ref{diffuse}(b,d)
demonstrates.  For example, if all four disks are far from
each other, then slight real-space motions change their relative
angles $\theta$ by small amounts.  Recall that cross products
such as Eqns.~\ref{cross} are related to $\sin \theta$ and dot
products such as Eqns.~\ref{dot} are related to $\cos \theta$,
so thus regions of the landscape corresponding to disks far
apart in real space have smaller changes in $\theta$ and therefore
have slower landscape motion.  Again, while this discussion
is considering diffusive dynamics, similar comments hold for
ballistic dynamics:  a constant real-space velocity leads to a
non-constant velocity through a landscape.

\section{Quantifying the entropic barrier}
\label{quantify}

Returning to the question of the power-law dependence
of the switching time scale $\tau$ on $\epsilon$ seen in
Fig.~\ref{lifetime}, we can understand this by recognizing it is an
entropic barrier.  Following Ref.~\cite{hunter12pre}, the barrier
can be quantified by counting the number of microstates $\Omega_t$
available at a transition.  A transition involves three collinear
disks:  the center disk is the one passing between the other
two, thus defining a swap, see for example Fig.~\ref{bigfig}(c,d).
If this line is along a diameter of the system, then the relative
positions of each of those three disks are described by just
three coordinates.  The length of the diameter is $6+2 \epsilon$,
but as each disk has a diameter of length 2, the amount of free
space is $2\epsilon$.  If one disk was confined to this much free
space, then $\Omega_t = 2\epsilon$.  For the three disks, while they
must share this free space, they each have $O(\epsilon)$ possible
positions and thus $\Omega_t \sim \epsilon^3$ for the three of them;
this can be confirmed by an exact calculation \cite{hunter12pre}.
However, transitions can occur when the disks are along a line other
than the diameter, so long as that line is at least of length $6$.
The position of that line has $O(\epsilon^{1/2})$ possibilities,
giving $\Omega_t \sim \epsilon^{7/2}$ for three disks to make a
transition \cite{hunter12pre}.  The fourth disk, which is not
as involved in the transition, nonetheless needs to be out of
the center of the system:  the number of microstates corresponding
to this extra degree of freedom is also proportional to $\epsilon$,
leading to the overall $\Omega_t \sim \epsilon^{9/2}$.

Compared to this transition state, the number of microstates
$\Omega_e$
associated with the equilibrium states is quite large, and
essentially independent of $\epsilon$ when $\epsilon \ll 1$.
Therefore growth of the entropic barrier as $\epsilon \rightarrow
0$ is determined by the $\epsilon$ dependence of $\Omega_t$
(related to the transition state).
This argument then suggests an entropic barrier that grows as
$F_{\rm B}/k_B T = -S_{\rm B}T / k_B T
\sim -\ln \Omega_t \sim \ln \epsilon^{-9/2}$.
In other words, the system has to find one of the rare transition
microstates counted by $\Omega$ as opposed to being in the
many microstates associated with a common configuration.
The scarcity of the transition microstates as $\epsilon$ becomes
small is what increases the entropic barrier, and thus slows
down the transition.  There is also a time scale $\tau_0$ for
attempts to cross the barrier, such that $\tau = \tau_0
\exp(F_B/k_B T) \sim \tau_0 \epsilon^{-9/2}$.  Figure
\ref{lifetime} shows this relation holds as $\epsilon \rightarrow
0$.  Note that this argument of counting the microstates does {\it
not} depend on defining a free energy landscape.  Rather, this is
a direct counting of microstates in the original 8 dimensional
state space, and thus does not have any of the arbitrariness of
defining new coordinates.

\section{Conclusions}
\label{conclusions}

We have presented a simple model system comprised of four disks
moving in a small region.  This system can be
described by several different free energy landscapes, with
greater or lesser success.  The spherical representation shown in
Fig.~\ref{spherespace} has the advantage of emphasizing the
symmetry of the landscape and the existence of six unique local
minima.  However, it has the drawback of requiring a 3D
representation, and thus is slightly harder to depict on the
printed page.  The simpler triangular
landscape of Fig.~\ref{diffuse}(a) collapses the six minima into
three, with the gained advantage of a purely 2D representation.
A different version of this triangular landscape, shown in
Fig.~\ref{diffuse}(c), has the disadvantage that the apparent
free energy barrier height is not as useful for determining the
transition rate between states.  These three landscapes
illustrate the main points we have made about free energy
landscapes:  (a) a system does not have ``the'' free energy
landscape, but rather multiple free energy landscapes can be defined
for a given system; (b) different free energy landscapes have
different apparent barrier heights; (c) the different
apparent barrier heights are compensated for by different
effective diffusivity rates on different landscapes, 
such that the transition rate between states is independent of
choice of free energy landscape description.  A related point is
that the effective diffusivity rate depends on the location in the
free energy landscape.  For well-chosen free energy landscapes
and in the limit of small system size, the transition time scale
between states has an Arrhenius scaling depending on the free
energy barrier height.  For simulations with ballistic dynamics,
the conclusions about diffusivity map smoothly to conclusions
about speed of trajectories through the different landscapes.

One additional point can be made by comparing this four disk model
with an earlier three disk model \cite{hunter12pre}.  The earlier
model has free energy landscapes describable by only one coordinate,
for example $\vec{v}_{12} \times \vec{v}_{13}$ (compare with our
Eqn.~\ref{cross}).  By adding one disk, we need to add at least
one coordinate in a useful free energy landscape description.
Clearly as we increase the number of disks (or consider spheres
moving in three dimensions) we will need more coordinates for a free
energy landscape description.  It seems likely that the number of
needed coordinates will scale as the number of particles $N$ for
large $N$, but exactly how this scaling should behave for large $N$
is unclear.  Nonetheless, it suggests that one can imagine that a
free energy landscape for $N \gg 1$ can be described by some space
with a dimensionality lower than the original coordinate space,
and the landscape will be highly symmetric albeit in some number
of dimensions hard to visualize.  It is plausible that explicitly
constructed free energy landscapes for large systems may be of
limited use given that they are still high-dimensional,
as is the original potential energy landscape.
Nonetheless we note that often authors do think about free
energy landscapes for hard particle systems ({\it e.g.},
\cite{bowles06,ashwin09,carlsson12,holmescerfon13,charbonneau14})
so it is encouraging to think that such landscapes could, at least
hypothetically, be constructed in a manner such as we have done
in this work.

A final comment is that if the particles in a system are not hard,
but interact with some interaction potential, then the potential
energy term $U$ contributes also to the free energy.  This situation
is considered elsewhere in the context of the earlier three disk
model \cite{du16}, which found that the entropic and energetic
contributions to the free energy landscape are often comparable.
That is, transitions can require both a thermal fluctuation that
allows particles to interact more strongly and increase $U$, and
also that particles find a rare, low entropy state.  Nonetheless,
the main points listed above for free energy landscapes will still
be true for situations with nontrivial potential energy.

We thank G.~L.~Hunter for helpful discussions.  This
work was supported by the National Science Foundation
(CBET-1336401 and CBET-1804186).


\begin{thebibliography}{10}
\newcommand{\enquote}[1]{``#1''}

\bibitem{goldstein69}
M.~Goldstein.
\newblock \enquote{Viscous liquids and the glass transition: A potential energy
  barrier picture.}
\newblock \emph{J. Chem. Phys.}, \textbf{51}, 3728--3739 (1969).

\bibitem{sciortino05}
F.~Sciortino \& P.~Tartaglia.
\newblock \enquote{Glassy colloidal systems.}
\newblock \emph{Adv. Phys.}, \textbf{54}, 471--524 (2005).

\bibitem{desouza08}
V.~K. de~Souza \& D.~J. Wales.
\newblock \enquote{Energy landscapes for diffusion: Analysis of cage-breaking
  processes.}
\newblock \emph{J. Chem. Phys.}, \textbf{129}, 164507 (2008).

\bibitem{bryngelson95}
J.~D. Bryngelson, J.~N. Onuchic, N.~D. Socci, \& P.~G. Wolynes.
\newblock \enquote{Funnels, pathways, and the energy landscape of protein
  folding: {A} synthesis.}
\newblock \emph{Proteins: Structure, Function, and Bioinformatics},
  \textbf{21}, 167--195 (1995).

\bibitem{berry97}
R.~S. Berry, N.~Elmaci, J.~P. Rose, \& B.~Vekhter.
\newblock \enquote{Linking topography of its potential surface with the
  dynamics of folding of a protein model.}
\newblock \emph{Proc. Nat. Acad. Sci.}, \textbf{94}, 9520--9524 (1997).

\bibitem{joseph17}
J.~A. Joseph, K.~R{\"o}der, D.~Chakraborty, R.~G. Mantell, \& D.~J. Wales.
\newblock \enquote{Exploring biomolecular energy landscapes.}
\newblock \emph{Chem. Comm.}, \textbf{53}, 6974--6988 (2017).

\bibitem{marcelin1914}
R.~Marcelin.
\newblock \enquote{Expression des vitesses de transformation des syst{\`e}mes
  physico-chimiques en fonction de l'affinit{\'e}.}
\newblock \emph{Compt. Rend. Hebd. Seances Acad. Sci.}, \textbf{158}, 116--118
  (1914).

\bibitem{awasthi19}
S.~Awasthi \& N.~N. Nair.
\newblock \enquote{Exploring high-dimensional free energy landscapes of
  chemical reactions.}
\newblock \emph{WIREs Comput. Mol. Sci.}, \textbf{9}, e1398 (2019).

\bibitem{krzakala07}
F.~Krzakala \& J.~Kurchan.
\newblock \enquote{Landscape analysis of constraint satisfaction problems.}
\newblock \emph{Physical Review E}, \textbf{76}, 021122 (2007).

\bibitem{ballard17}
A.~J. Ballard, R.~Das, S.~Martiniani, D.~Mehta, L.~Sagun, J.~D. Stevenson, \&
  D.~J. Wales.
\newblock \enquote{Energy landscapes for machine learning.}
\newblock \emph{Phys. Chem. Chem. Phys.}, \textbf{19}, 12585--12603 (2017).

\bibitem{laidler85}
K.~J. Laidler.
\newblock \enquote{Rene {Marcelin} (1885-1914), a short-lived genius of
  chemical kinetics.}
\newblock \emph{J. Chem. Educ.}, \textbf{62}, 1012 (1985).

\bibitem{stillinger84}
F.~H. Stillinger \& T.~A. Weber.
\newblock \enquote{Packing structures and transitions in liquids and solids.}
\newblock \emph{Science}, \textbf{225}, 983--989 (1984).

\bibitem{stillinger88}
F.~H. Stillinger.
\newblock \enquote{Supercooled liquids, glass transitions, and the {Kauzmann}
  paradox.}
\newblock \emph{J. Chem. Phys.}, \textbf{88}, 7818--7825 (1988).

\bibitem{stillinger95}
F.~H. Stillinger.
\newblock \enquote{A topographic view of supercooled liquids and glass
  formation.}
\newblock \emph{Science}, \textbf{267}, 1935--1939 (1995).

\bibitem{alder57}
B.~J. Alder \& T.~E. Wainwright.
\newblock \enquote{Phase transition for a hard sphere system.}
\newblock \emph{J. Chem. Phys.}, \textbf{27}, 1208--1209 (1957).

\bibitem{wood57}
W.~W. Wood \& J.~D. Jacobson.
\newblock \enquote{Preliminary results from a recalculation of the {M}onte
  {C}arlo equation of state of hard spheres.}
\newblock \emph{J. Chem. Phys.}, \textbf{27}, 1207--1208 (1957).

\bibitem{bernal64}
J.~D. Bernal.
\newblock \enquote{The {B}akerian lecture, 1962. {T}he structure of liquids.}
\newblock \emph{Proc. Roy. Soc. London. Series A}, \textbf{280}, 299--322
  (1964).

\bibitem{widom67}
B.~Widom.
\newblock \enquote{Intermolecular forces and the nature of the liquid state.}
\newblock \emph{Science}, \textbf{157}, 375--382 (1967).

\bibitem{pusey86}
P.~N. Pusey \& W.~van Megen.
\newblock \enquote{Phase behaviour of concentrated suspensions of nearly hard
  colloidal spheres.}
\newblock \emph{Nature}, \textbf{320}, 340--342 (1986).

\bibitem{zhou91}
H.~X. Zhou \& R.~Zwanzig.
\newblock \enquote{A rate process with an entropy barrier.}
\newblock \emph{J. Chem. Phys.}, \textbf{94}, 6147--6152 (1991).

\bibitem{hunter12pre}
G.~L. Hunter \& E.~R. Weeks.
\newblock \enquote{Free-energy landscape for cage breaking of three hard
  disks.}
\newblock \emph{Phys. Rev. E}, \textbf{85}, 031504 (2012).

\bibitem{hill76}
T.~L. Hill \& E.~Eisenberg.
\newblock \enquote{Reaction free energy surfaces in myosin-actin-{ATP}
  systems.}
\newblock \emph{Biochemistry}, \textbf{15}, 1629--1635 (1976).

\bibitem{soukoulis82}
C.~M. Soukoulis, K.~Levin, \& G.~S. Grest.
\newblock \enquote{Reversibility and {Irreversibility} in {Spin}-{Glasses}:
  {The} {Free}-{Energy} {Surface}.}
\newblock \emph{Phys. Rev. Lett.}, \textbf{48}, 1756--1759 (1982).

\bibitem{paine85}
G.~H. Paine \& H.~A. Scheraga.
\newblock \enquote{Prediction of the native conformation of a polypeptide by a
  statistical-mechanical procedure. {I}. {Backbone} structure of enkephalin.}
\newblock \emph{Biopolymers}, \textbf{24}, 1391--1436 (1985).

\bibitem{fontana91}
W.~Fontana, T.~Griesmacher, W.~Schnabl, P.~F. Stadler, \& P.~Schuster.
\newblock \enquote{Statistics of landscapes based on free energies, replication
  and degradation rate constants of {RNA} secondary structures.}
\newblock \emph{Monatshefte f{\"u}r Chemie}, \textbf{122}, 795--819 (1991).

\bibitem{wales03}
D.~Wales.
\newblock \emph{Energy {Landscapes}: {Applications} to {Clusters}, {Biomolecule
  s} and {Glasses}}.
\newblock Cambridge {Molecular} {Science} (Cambridge University Press,
  Cambridge) (2004).
\newblock ISBN 978-0-521-81415-7.

\bibitem{wales06}
D.~J. Wales \& T.~V. Bogdan.
\newblock \enquote{Potential energy and free energy landscapes.}
\newblock \emph{J. Phys. Chem. B}, \textbf{110}, 20765--20776 (2006).

\bibitem{giarritta94}
S.~P. Giarritta \& P.~V. Giaquinta.
\newblock \enquote{Statistical geometry of four calottes on a sphere.}
\newblock \emph{J. Stat. Phys.}, \textbf{75}, 1093--1118 (1994).

\bibitem{speedy94}
R.~J. Speedy.
\newblock \enquote{Two disks in a box.}
\newblock \emph{Physica A: Statistical Mechanics and its Applications},
  \textbf{210}, 341--351 (1994).

\bibitem{bowles99}
R.~K. Bowles \& R.~J. Speedy.
\newblock \enquote{Five discs in a box.}
\newblock \emph{Physica A}, \textbf{262}, 76--87 (1999).

\bibitem{bowles06}
R.~K. Bowles \& I.~Saika-Voivod.
\newblock \enquote{Landscapes, dynamic heterogeneity, and kinetic facilitation
  in a simple off-lattice model.}
\newblock \emph{Phys. Rev. E}, \textbf{73}, 011503 (2006).

\bibitem{ashwin09}
S.~S. Ashwin \& R.~K. Bowles.
\newblock \enquote{Complete jamming landscape of confined hard discs.}
\newblock \emph{Phys. Rev. Lett.}, \textbf{102}, 235701 (2009).

\bibitem{carlsson12}
G.~Carlsson, J.~Gorham, M.~Kahle, \& J.~Mason.
\newblock \enquote{Computational topology for configuration spaces of hard
  disks.}
\newblock \emph{Phys. Rev. E}, \textbf{85}, 011303 (2012).

\bibitem{barnettjones13}
M.~Barnett-Jones, P.~A. Dickinson, M.~J. Godfrey, T.~Grundy, \& M.~A. Moore.
\newblock \enquote{Transition state theory and the dynamics of hard disks.}
\newblock \emph{Phys. Rev. E}, \textbf{88}, 052132 (2013).

\bibitem{hinow14}
P.~Hinow.
\newblock \enquote{A nonsmooth program for jamming hard spheres.}
\newblock \emph{Optim. Lett.}, \textbf{8}, 13--33 (2014).

\bibitem{godfrey14}
M.~J. Godfrey \& M.~A. Moore.
\newblock \enquote{Static and dynamical properties of a hard-disk fluid
  confined to a narrow channel.}
\newblock \emph{Phys. Rev. E}, \textbf{89}, 032111 (2014).

\bibitem{godfrey15}
M.~J. Godfrey \& M.~A. Moore.
\newblock \enquote{Understanding the ideal glass transition: Lessons from an
  equilibrium study of hard disks in a channel.}
\newblock \emph{Phys. Rev. E}, \textbf{91} (2015).

\bibitem{robinson16}
J.~F. Robinson, M.~J. Godfrey, \& M.~A. Moore.
\newblock \enquote{Glasslike behavior of a hard-disk fluid confined to a narrow
  channel.}
\newblock \emph{Phys. Rev. E}, \textbf{93} (2016).

\bibitem{ediger12}
M.~D. Ediger \& P.~Harrowell.
\newblock \enquote{Perspective: Supercooled liquids and glasses.}
\newblock \emph{J. Chem. Phys.}, \textbf{137}, 080901 (2012).

\bibitem{biroli13}
G.~Biroli \& J.~P. Garrahan.
\newblock \enquote{Perspective: The glass transition.}
\newblock \emph{J. Chem. Phys.}, \textbf{138}, 12A301 (2013).

\bibitem{donati98}
C.~Donati, J.~F. Douglas, W.~Kob, S.~J. Plimpton, P.~H. Poole, \& S.~C.
  Glotzer.
\newblock \enquote{Stringlike cooperative motion in a supercooled liquid.}
\newblock \emph{Phys. Rev. Lett.}, \textbf{80}, 2338--2341 (1998).

\bibitem{doliwa00}
B.~Doliwa \& A.~Heuer.
\newblock \enquote{Cooperativity and spatial correlations near the glass
  transition: Computer simulation results for hard spheres and disks.}
\newblock \emph{Phys. Rev. E}, \textbf{61}, 6898--6908 (2000).

\bibitem{frenkel13}
D.~Frenkel.
\newblock \enquote{Simulations: The dark side.}
\newblock \emph{Euro. Phys. J. Plus}, \textbf{128}, 1--21 (2013).

\bibitem{holmescerfon13}
M.~Holmes-Cerfon, S.~J. Gortler, \& M.~P. Brenner.
\newblock \enquote{A geometrical approach to computing free-energy landscapes
  from short-ranged potentials.}
\newblock \emph{Proc. Nat. Acad. Sci.}, \textbf{110}, E5--E14 (2013).

\bibitem{charbonneau14}
P.~Charbonneau, J.~Kurchan, G.~Parisi, P.~Urbani, \& F.~Zamponi.
\newblock \enquote{Fractal free energy landscapes in structural glasses.}
\newblock \emph{Nature Comm.}, \textbf{5}, 4725 (2014).

\bibitem{du16}
X.~Du \& E.~R. Weeks.
\newblock \enquote{Energy barriers, entropy barriers, and {non-Arrhenius}
  behavior in a minimal glassy model.}
\newblock \emph{Phys. Rev. E}, \textbf{93}, 062613 (2016).

\end{thebibliography}
\end{document}